%% file: eprint.tex
\documentclass[10pt, paper=a4, UKenglish]{article}
\usepackage{graphicx}
\usepackage{amsmath}
\usepackage{placeins}
\def\Title#1{\begin{center} {\Large #1 } \end{center}}
\def\Author#1{\begin{center}{ \sc #1} \end{center}}
\def\Address#1{\begin{center}{ \it #1} \end{center}}

\newcommand\pubblock{\rightline{\begin{tabular}{l} Proceedings of the CTD 2025\\ \pubnumber\\
         \pubdate  \end{tabular}}}

\newenvironment{Abstract}{%
    \begin{center}\large ABSTRACT\end{center}\bigskip
    \begin{quotation}\begin{center}\large
}{%
    \end{center}\end{quotation}
}

\newenvironment{Presented}{%
    \begin{center}\large PRESENTED AT\end{center}\bigskip
    \begin{quotation}\begin{center}\large
}{%
    \end{center}\end{quotation}
}

\def\Acknowledgements{\bigskip  \bigskip \begin{center} \begin{large}
      \bf ACKNOWLEDGEMENTS \end{large}\end{center}}

\input econfmacros.tex

\usepackage{color}
\usepackage{lineno}
\usepackage{subfig}
\usepackage{hyperref}

\usepackage[backend=biber,style=numeric,sorting=none,maxnames=5, minnames=5]{biblatex}
\def\affiliation{On behalf of the\\
ATLAS Collaboration, \\
CERN, Geneva, Switzerland}

\newcommand\pubnumber{PROC-CTD2025-81}

\newcommand\pubdate{\today}

\newcommand{\conference}{Connecting the Dots Workshop (CTD 2025)\\
November 10-14, 2025}

\usepackage{fancyhdr}
\definecolor{mygrey}{RGB}{105,105,105}
\begin{document}



\large
\begin{titlepage}
\pubblock

\vfill
\Title{Vision Transformers and Graph Neural Networks for Charged Particle Tracking in the ATLAS Muon Spectrometer}
\vfill

\Author{Jonathan Renusch}
\Address{\affiliation}

\vfill

\begin{Abstract}
  The identification and reconstruction of charged particles, such as muons, is a main challenge for the physics program of the ATLAS experiment at the Large Hadron Collider. This task will become increasingly difficult with the start of the High-Luminosity LHC era after 2030, when the number of proton-proton collisions per bunch crossing will increase from 60 to up to 200. This elevated interaction density will also increase the occupancy within the ATLAS Muon Spectrometer, requiring more efficient and robust real-time data processing strategies within the experiment's trigger system, particularly the Event Filter. To address these algorithmic challenges, we present two machine-learning-based approaches. First, we target the problem of background-hit rejection in the Muon Spectrometer using Graph Neural Networks integrated into the non-ML baseline reconstruction chain, demonstrating a 15\% improvement in reconstruction speed (from 255\,ms to 217\,ms). Second, we present a proof-of-concept for end-to-end muon tracking using state-of-the-art Vision Transformer architectures, achieving ultra-fast approximate muon reconstruction in 2.3\,ms on consumer-grade GPUs at 98 \% tracking efficiency.
\end{Abstract}

\vfill

\begin{Presented}
\conference
\end{Presented}
\vfill
{\footnotesize \noindent © 2026 CERN for the benefit of the ATLAS Collaboration. Reproduction of this article or parts of it is allowed as specified in the CC-BY-4.0 license.}
\end{titlepage}
\def\thefootnote{\fnsymbol{footnote}}
\setcounter{footnote}{0}
%

\normalsize 


\section{Introduction}
\label{intro}

Modern high-energy collider physics aims to probe the fundamental constituents of matter via highly energetic proton-proton collisions. The Large Hadron Collider (LHC) at CERN serves as the primary instrument for this exploration, providing four main interaction points for particle collisions. One of these points hosts the ATLAS experiment \cite{ATLAS_paper}, the largest collision experiment on earth, which serves as a general-purpose detector designed for the discovery of physics beyond the Standard Model (BSM) and the precision characterization of known physical processes. Within this experimental framework, charged particle reconstruction is fundamental to almost all physics workflows. In both the search for new physics and the precision measurement of rare processes, the trajectories of charged particles are among the first signals extracted from the data. The reconstruction task can be broken down into three sub-problems. First, the number of particles produced during the collision event must be detected. Second, the energy deposits left by particles in the detector (referred to as ``hits'') must be assigned to the correct particle trajectories. Finally, any method must be able to estimate particle-specific properties such as transverse momentum, and azimuthal and polar angles.

This work is situated within the context of a future upgrade of the LHC, the High-Luminosity LHC (HL-LHC) \cite{Aberle:2749422}. The HL-LHC era, starting with the data-taking period known as Run 4, will present unprecedented challenges for particle reconstruction. The number of proton-proton collisions per bunch crossing, referred to as pileup ($\langle\mu\rangle$), is expected to increase significantly from approximately $\langle\mu\rangle=60$ in the current Run 3 to up to $\langle\mu\rangle=200$ during the HL-LHC operation. This dense environment necessitates robust and efficient algorithms, particularly for the experiment's real-time selection system, the Trigger. Muon reconstruction is a critical component of the second level of this system, known as the Event Filter (EF), where strict latency constraints require highly optimized processing to select interesting physics events from the 1\,MHz input rate for permanent storage.

Compared to Inner Tracker and calorimeter-like detector geometries \cite{GNN4ITk,flavourTagging,ATLAS_paper}, Transformer and Graph Neural Network (GNN) based methods are relatively unexplored in the context of Muon Spectrometer reconstruction. Despite collecting approximately one order of magnitude fewer hits per collision event ($O(10\,\text{k})$ vs. $O(100\,\text{k})$) compared to the aforementioned Inner Tracker, the Muon Spectrometer poses two particular challenges for charged particle tracking \cite{ATLAS_paper}. First, it features an extremely low signal-to-noise ratio; in some of the ATLAS simulation datasets investigated in this study, the signal-to-noise ratio is as low as 0.6\%, and in certain detector regions such as the New Small Wheel (NSW), this ratio can be significantly lower. Second, the Muon Spectrometer complex integrates five different sub-detector technologies\footnote{Monitored Drift Tubes (MDT), Resistive Plate Chambers (RPC), Thin Gap Chambers (TGC), Micromegas (MM), and small-strip Thin Gap Chambers (sTGC).} with varying readout speeds, underlying detection principles, and spatial precisions. Furthermore, the 0.5\,T magnetic field is inhomogeneous due to the metallic scaffolding and wires within the Muon Spectrometer, which spans a $25 \times 46$\,m cylindrical volume. From a Machine Learning (ML) perspective, these constraints require a multi-task ML architecture that is able to process a multi-modal time series derived from heterogeneous spatial geometries.

This work summarizes two different philosophies for addressing the described problem.
The first approach investigates the use of GNNs to improve the existing baseline reconstruction algorithm by rejecting background hits prior to track finding within local regions of the Muon Spectrometer. Inspired by a publication on the TrackML dataset \cite{hepattn, Amrouche_2019}, the second idea extends the local approach by proposing a global Transformer-based filtering stage followed by a state-of-the-art image segmentation model to solve the combinatorial problems of particle detection and hit-to-track assignment in a purely machine-learning-based manner.

Despite following different design philosophies, both methods share the initial goal of removing background hits before the main reconstruction stage. In general, background hits in the Muon Spectrometer originate from various sources, including secondary particles produced via interactions with detector material, cosmic rays, electronic noise, gamma radiation, and isotopic decays in the detector cavern. The presence of background hits can lead to incorrect track assignments, reduced efficiency in muon identification, and increased computational load during the matching process.

The baseline reconstruction algorithm executed on CPU hardware consists of sequential stages. First, a custom clustering algorithm assigns hits in the detector to clusters, referred to as ``Muon Buckets'' \cite{DiCroce:2946578}. Specifically, this algorithm groups hits if they are within 30\,cm of each other along the longitudinal direction. Buckets are divided into two separate clusters if their combined length would exceed a total length of 2\,m. Every signal and background hit is assigned to a bucket. This scheme was developed independently of ML-based hit filtering to constrain the search space of subsequent pattern recognition algorithms and to facilitate efficient data management. Following clustering, a pattern recognition algorithm identifies potential muon track candidates by searching for spatially aligned hits across different Muon Spectrometer layers using a Hough transform within each Muon Bucket. Subsequently, a Maximum Likelihood-based $\chi^2$ track fit is performed to estimate the direction of local track segments within each bucket. The coarse direction estimates provided by these segments are then used to solve the global combinatorial problem of hit-to-track assignment across the entire spectrometer. Finally, the global track parameters are determined through a global $\chi^2$ track fit on each track. The entire baseline reconstruction chain takes roughly 255\,ms of runtime per event at $\langle\mu\rangle=200$ when executed on a single CPU thread on an AMD EPYC 9654 CPU; this benchmark was obtained using a preliminary setup and is expected to improve with further algorithmic optimisations.

\section{Graph Neural Networks for Background Rejection}
\label{gnn}

GNNs are a class of machine learning models designed to operate on graph-structured data. They have demonstrated significant success in domains with strong inductive biases or where the data scale or structure makes full attention matrices, such as those used in standard Transformers, computationally infeasible. This first scenario is particularly relevant when modeling physical systems where local connectivity is a defining feature such as weather models \cite{graphcast}, molecular structures \cite{AlphaFold3} and High Energy Physics collision events \cite{GNN4ITk}.

The models propagate information through the graph by implementing message-passing algorithms. These algorithms iteratively update node and edge representations based on their local neighborhood, effectively mimicking the flow of information (or matter) in physical systems. Given the sparse geometry of hits in the Muon Spectrometer, GNNs provide a natural starting point for the classification of signal against background noise. 

Although low-level studies for hit filtering with individual hits as nodes in the graph have been attempted, further investigations showed that this is not a computationally viable approach. Therefore, it was decided to build graphs from higher-order clusters of hits (Muon Buckets) in the interest of computational efficiency. The graph is constructed dynamically using the spatial coordinates of the Muon Bucket \cite{DiCroce:2946578}. To ensure data quality at the node level, only buckets meeting specific thresholds—such as containing hits across more than two detector layers—are included as valid nodes. Connectivity between these nodes is restricted to buckets in identical or adjacent detector sectors that satisfy the following spatial proximity criteria:
\begin{equation}
    |\Delta z| < 15 000\,\text{mm} \quad \text{and} \quad \sqrt{(\Delta x)^2 + (\Delta y)^2} < 6 800\,\text{mm},
\end{equation}
where $(\Delta x, \Delta y, \Delta z)$ represent the spatial separations between the bucket centres. The graphs do not have self-connections. This edge-building strategy ensures that local spatial correlations within the Muon Spectrometer are captured while maintaining a sparse graph structure facilitating computationally efficient message passing.

\begin{figure}[!htb]
  \centering
  \includegraphics[width=\linewidth, trim=192pt 162pt 192pt 162pt, clip]{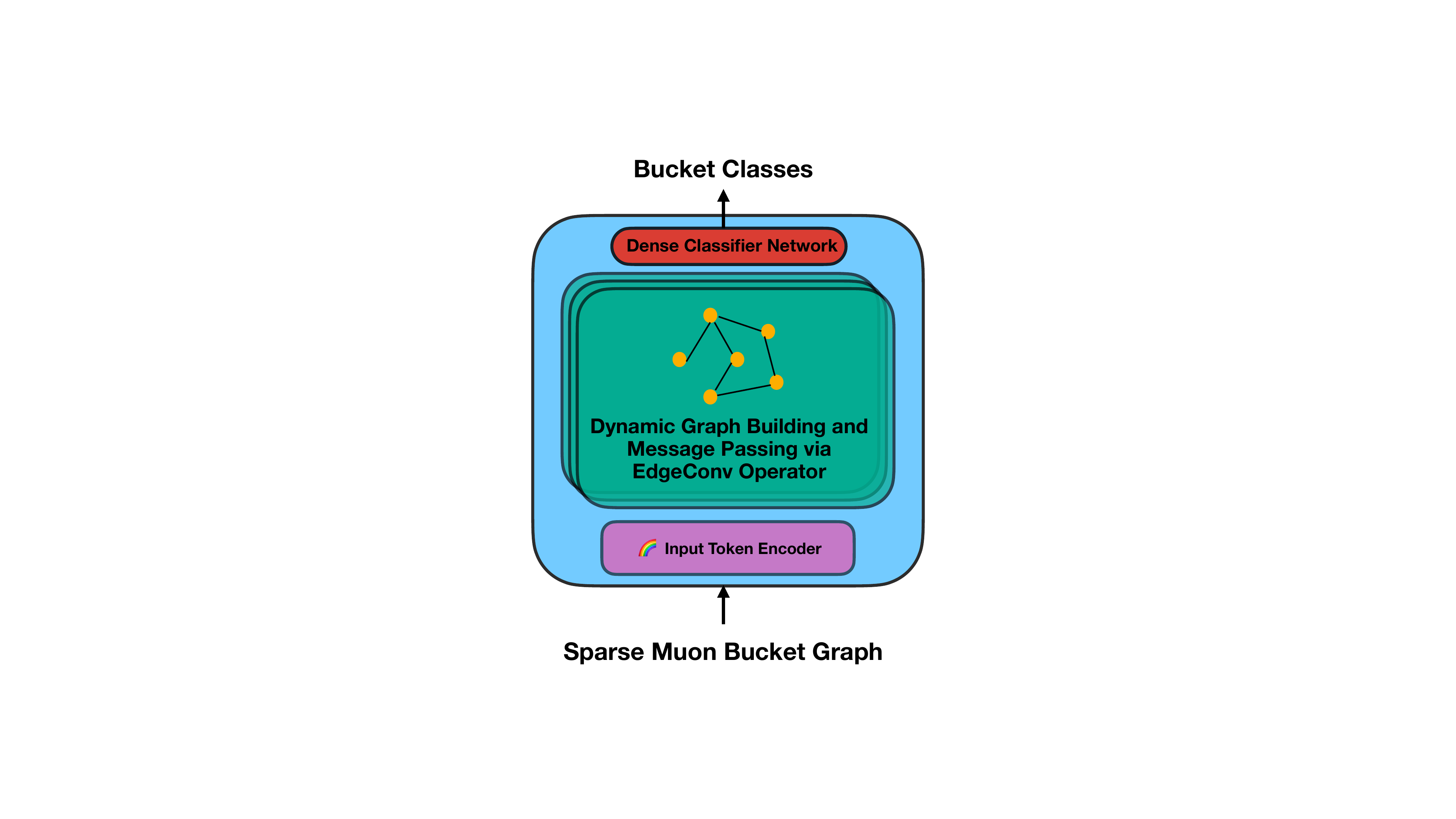}
  \caption{Architecture of the EdgeConv-based GNN used for background hit rejection. The model operates on graphs constructed from Muon Buckets to achieve high computational efficiency.}
  \label{fig:edgeconv}
\end{figure}

The GNN deployed into the ATLAS Athena framework \cite{Athena} for hit filtering is primarily based on the EdgeConv \cite{EdgeConv} architecture, as illustrated in Figure~\ref{fig:edgeconv}. A significant technical challenge was the integration of PyTorch Geometric with the Open Neural Network Exchange (ONNX) format for inference within the C++ ATLAS reconstruction framework. In particular, the EdgeConv operator, which relies on custom CUDA kernels and dynamic graph indexing implemented in the Pytorch Geometric library \cite{PyGeometric}, is not natively supported in ONNX, necessitating a custom implementation for deployment.

The study uses full ATLAS simulation datasets with Run 4 Geometry and mainly Z boson decays to muons ($Z \to \mu\mu$) and different pile-ups of $\langle\mu\rangle=60 \ \text{to} \ 200$. The model is trained to differentiate between Muon Buckets containing one segment, multiple segments or background noise. For further specific details regarding the model architecture, training procedure, and more performance benchmarks, please refer to the ATLAS Note published \cite{DiCroce:2946578}. Code on the explored architectures is available at the \href{https://gitlab.cern.ch/atlas-nextgen/work-package-2.5/StationHitClassifier}{\textbf{StationHitClassifier}} repository.

\subsection{GNN Results}
\begin{figure}[!htb]
  \centering
  \includegraphics[width=0.95\linewidth]{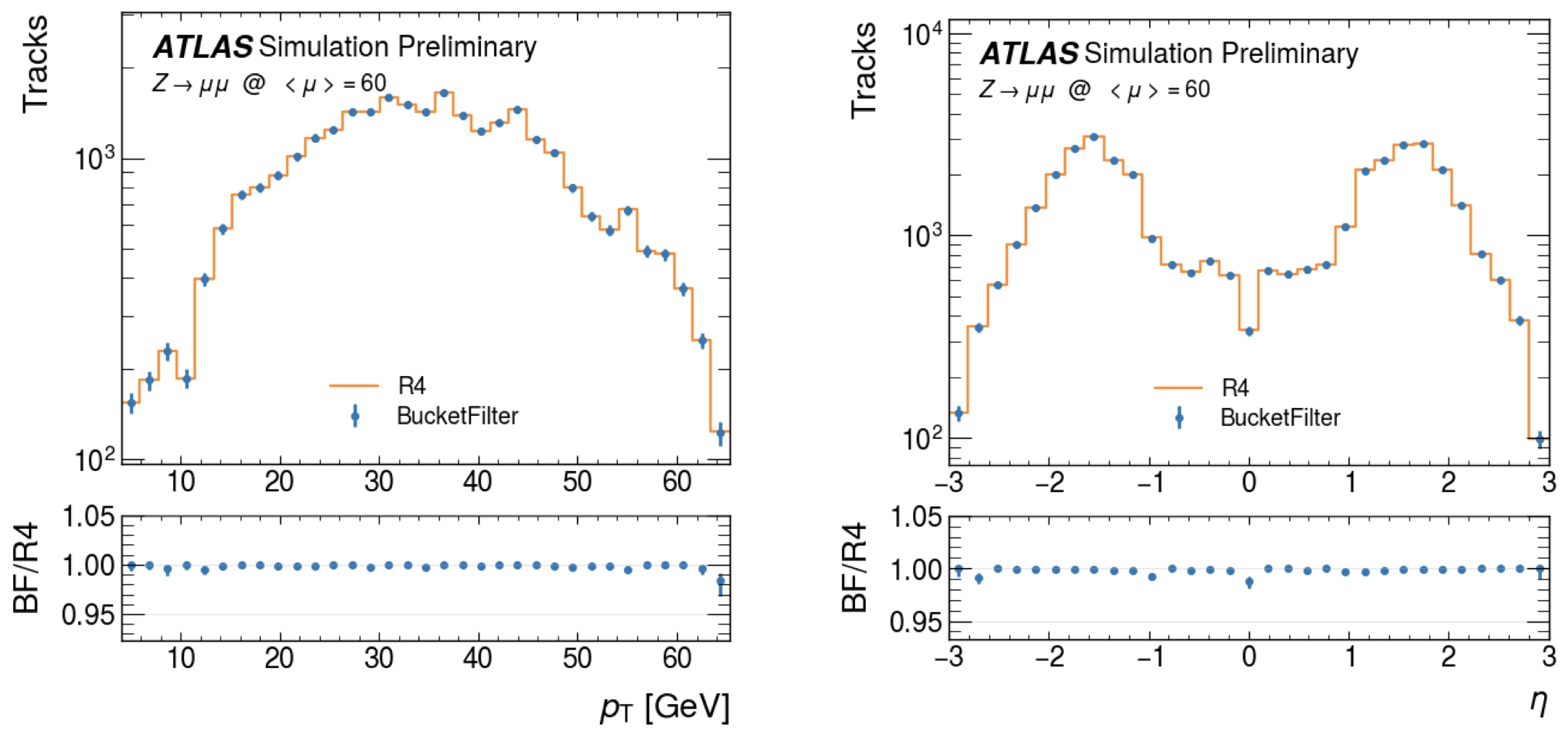}
  \caption{Comparison of reconstructed muon kinematic distributions for the standard R4 reconstruction (orange) and the chain equipped with the GNN-based Bucket Filter (blue), matched to generator-level muons \cite{GGN_MuonBucketFiltering}. The bottom panels show the ratio between the two methods, indicating no significant loss in performance.}
  \label{fig:ratios}
\end{figure}

\begin{figure}[!htb]
  \centering
  \includegraphics[width=0.75\linewidth]{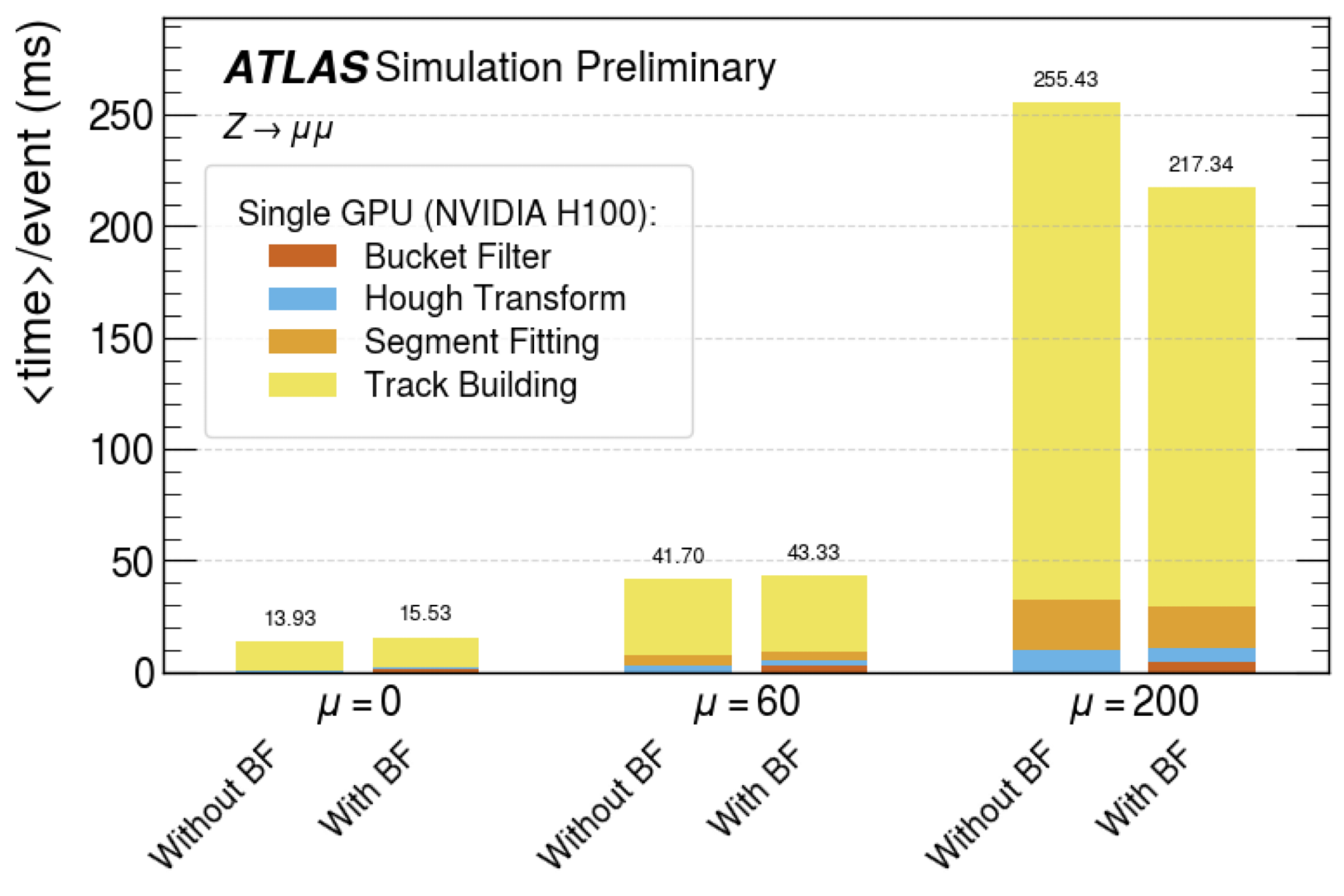}
  \caption{Average per-event execution time of the muon reconstruction chain for $Z \to \mu\mu$ events across various pileup levels \cite{GGN_MuonBucketFiltering}. The application of the Bucket Filter yields approx. a 15\% improvement in total processing speed at $\langle\mu\rangle=200$.}
  \label{fig:GNN_speed}
\end{figure}

The performance of the GNN-based Bucket Filter was evaluated on simulated $Z \to \mu\mu$ events across various pileup levels. We find that the model achieves a background bucket rejection rate of approx. 97\% at $\langle\mu\rangle=60$. As shown in Figure~\ref{fig:GNN_speed}, this reduction in input data for the subsequent pattern recognition stages translates into a significant decrease in total execution time when running the Athena baseline algorithm, as outlined in the introduction. For high-occupancy events at $\langle\mu\rangle=200$, the integration of the Bucket Filter reduces the average per-event reconstruction time by 15\% when evaluated using an NVIDIA H100 GPU and AMD EPYC 9654 CPU. Crucially, the signal reconstruction efficiency and precision of the baseline algorithm are preserved across the entire kinematic range, as shown in Figure~\ref{fig:ratios}, demonstrating that the GNN integration does not introduce any negative bias into the physics performance.

\section{Vision Transformers for End-to-End Muon Reconstruction}
\label{vit}
The second approach explored in this work leverages modern developments in the attention algorithm via windowed Flash Attention and computer vision, specifically the Vision Transformer (ViT) architecture. By relying on attention mechanisms and state-of-the-art ViTs as the primary algorithmic choice, we benefit from the extensive research and architectural optimizations pioneered in the computer vision community, as well as broader hardware and software developments in attention algorithms. In particular, this work uses an adaptation of the Mask2Former architecture \cite{Mask2Former,hepattn}. Mask2Former models have established themselves as an ML-baseline for panoptic image segmentation\footnote{Panoptic image segmentation is the task of classifying every pixel into a category while simultaneously segmenting individual object instances from one another.}, which is highly applicable to this tracking task.

Typical ViTs consist of two main components: a pixel-level encoder that encodes patches of pixels into tokens and a transformer decoder. These encoders often rely on Convolutional Neural Networks (CNNs) or simpler attention-based transformers to extract low-level features from the input image. The transformer decoder then processes these features to generate high-level representations suitable for downstream tasks such as object detection or segmentation. Within the Mask2Former model, the decoder stage performs cross-attention between learnable queries and patch tokens. Additionally, each layer of the decoder predicts an attention mask that is reinjected into subsequent layers. This mask is calculated using a multilayer perceptron (MLP) on the queries, followed by a dot product with the patch tokens. The mask is then binarised using a sigmoid function at a specific threshold. For specific implementation details, please refer to Refs.~\cite{Mask2Former, MaskFormer,hepattn}. These attention masks are a core innovation of the MaskFormer architecture, allowing it to iteratively refine the segmentation boundaries of objects in each layer of the decoder while masking out background noise that has already been discarded as such in previous layers. Furthermore, the decoder performs self-attention between the queries in each layer to facilitate information flow between object candidates and mitigate duplicate reconstructions. Since the second iteration of the Mask2Former architecture, these object queries (track candidates) are implemented as learned vectors during the backpropagation.

\begin{figure}[!htb]
  \centering
  \includegraphics[width=0.8\linewidth]{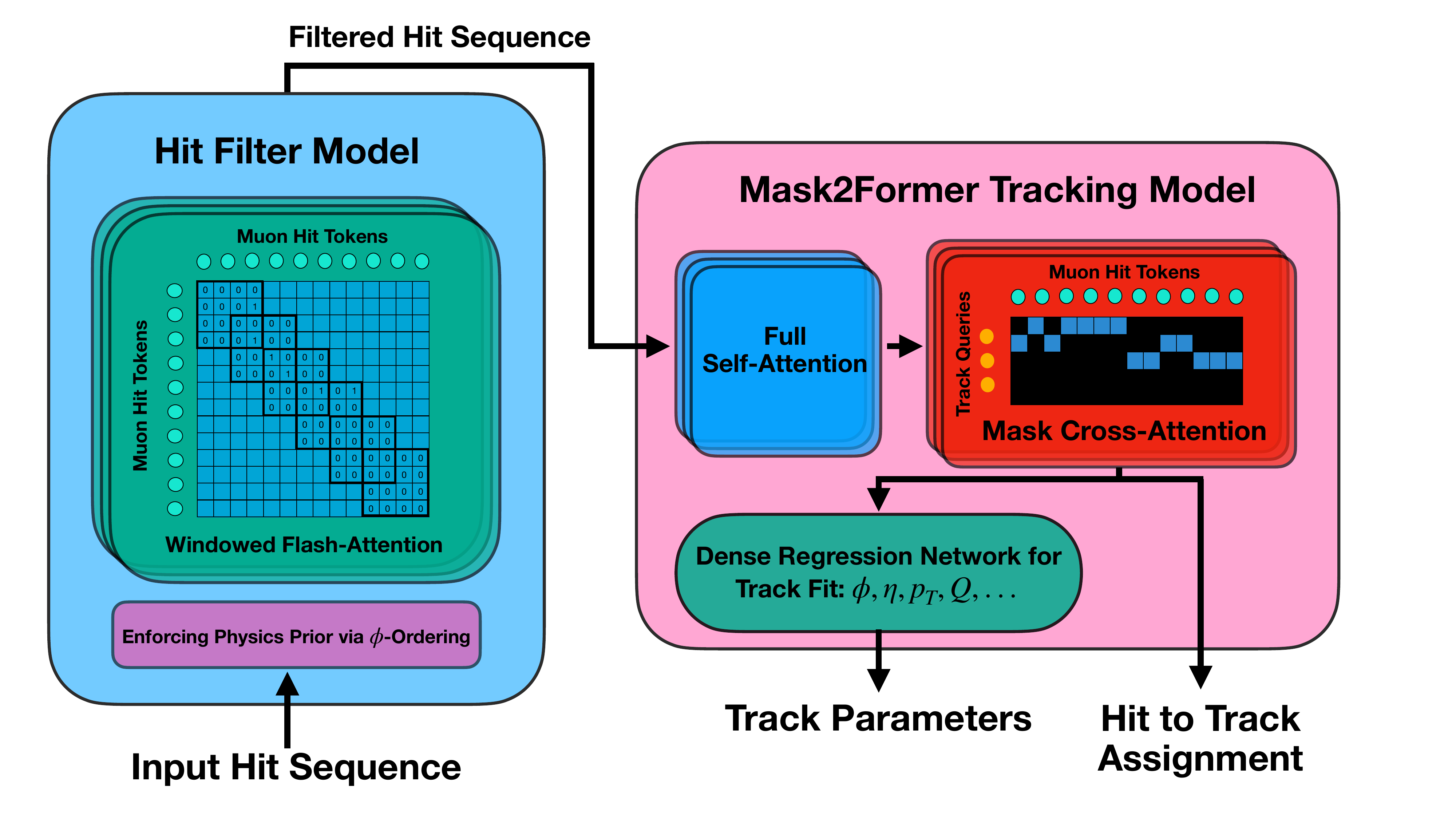}
  \caption{Schematic of the Transformer-based tracking architecture adapted from the Mask2Former model \cite{Mask2Former, hepattn}. The model treats individual hits as tokens, allowing the decoder to iteratively refine track candidates (queries) through mask-conditioned cross-attention.}
  \label{fig:MaskFormer}
\end{figure}

The High Energy Physics (HEP) tracking implementation introduced in Ref.~\cite{hepattn} adapts the core features of the Mask2Former architecture, including mask-conditioned cross-attention and the iterative refinement of object boundaries, but incorporates several domain-specific modifications: Instead of patch-level embeddings, the architecture treats individual detector hits as separate tokens. This approach avoids the need for complex upsampling strategies to produce per-pixel labels, which would be difficult to implement given the sparse and heterogeneous structure of the particle collision data. To maintain per-hit resolution against a high-noise background, the proposed pipeline instead incorporates a hit-filtering stage before the actual particle tracking stage. This filter performs binary classification to discriminate between signal and noise at the individual hit level. The model employs a physics-informed prior by sorting hits in the azimuthal angle $\phi$, assuming that only hits within a certain proximity in $\phi$ can originate from the same track. Based on this assumption, a windowed Flash Attention implementation \cite{FlashAttention3} is used, which scales $O(W \times N)$ in computational complexity, where $N$ is the sequence length and $W$ is the window size. This avoids the usual quadratic scaling behaviour of the full attention mechanism and presents one of the core innovations of this effort \cite{hepattn}. To allow hits to communicate across the $\pm\pi$ boundary, hits within half the window length are appended to each side of the sequence \cite{hepattn}. The overall architecture is illustrated in Figure~\ref{fig:MaskFormer}.

\subsection{ViT Simulation Dataset}
\begin{figure}[!htb]
  \centering
  \includegraphics[width=0.8\linewidth]{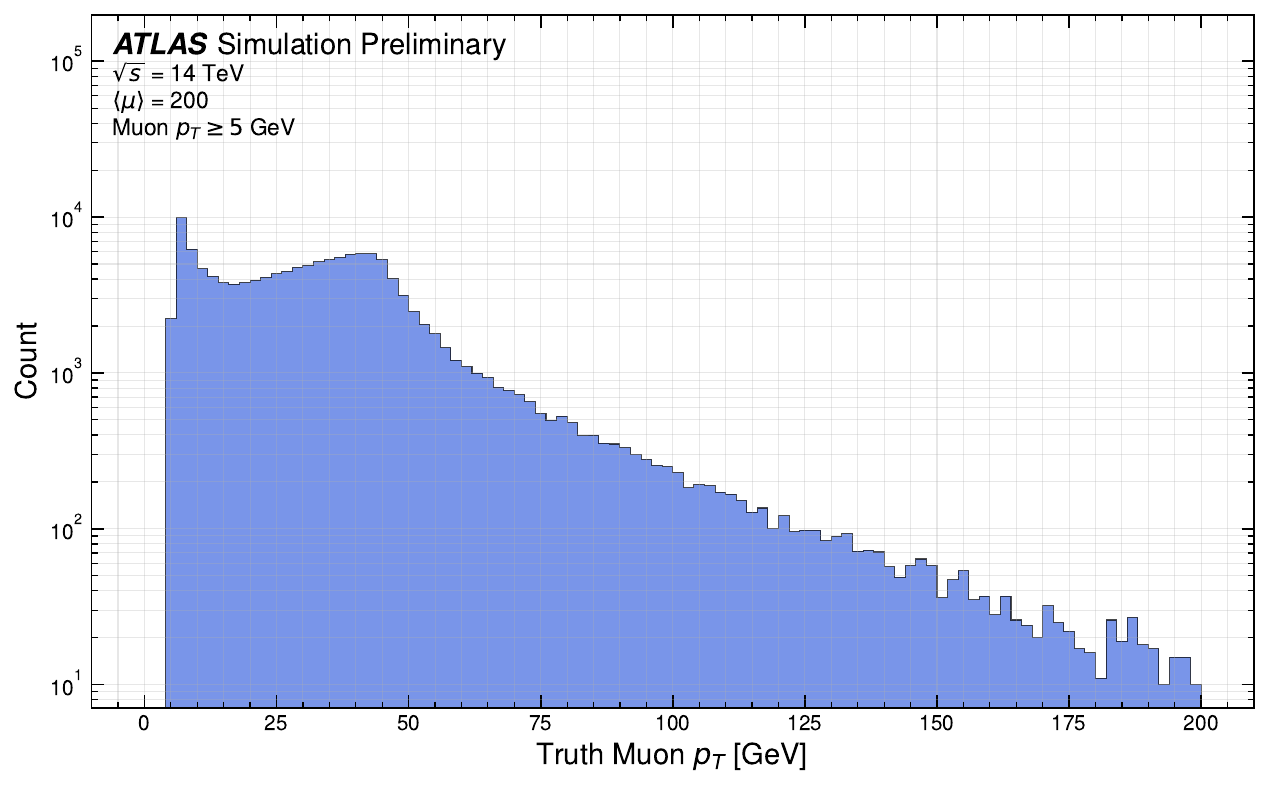}
  \caption{Transverse momentum distribution of signal muons in the testing dataset used for ViT-based tracking \cite{ATLAS_MDET_2025}. The dataset includes $J/\psi \to \mu\mu$, $t\bar{t}$, and $Z \to \mu\mu$ events simulated at HL-LHC conditions ($\langle\mu\rangle=200$).}
  \label{fig:data_pu200}
\end{figure}

This study utilizes full ATLAS simulations of the Muon Spectrometer with Run 4 geometry at $\sqrt{s} = 14$\,TeV. The dataset comprises a mixture of $Z \to \mu\mu$, $J/\psi \to \mu\mu$, and $t\bar{t}$ events at a pileup of $\langle\mu\rangle=200$. Target muons must have at least $p_{\text{T}} \ge 5$\,GeV of transverse momentum and three true hits to be considered for reconstruction; particles below this threshold are treated as noise by the filtering stage. The kinematic properties of this testing dataset are provided in Figure~\ref{fig:data_pu200}, while a representative event display is shown in Figure~\ref{fig:event_display}. 

The model was trained on about 1.4 million events and validated on 77,000 events. The independent testing dataset used to produce the results shown in this study consists of 85,000 events. Each event contains one or two muon tracks and an average occupancy of approx. 6,900 total hits with an initial signal-to-noise ratio of approx. 0.6\%. For tracks to be considered as targets for the ViT-based tracking stage, they must retain at least three associated hits in the Muon Spectrometer after the hit filtering stage.

\subsection{ViT Implementation Details}
The hit filtering and tracking stages of the proposed architecture are trained in separate steps. The filtering stage utilizes a windowed Flash Attention mechanism \cite{FlashAttention3} by Tri Dao with a window size of 512, consisting of 8 layers with 8 attention heads, RMS pre-normalization, SwiGLU activation and a hidden dimension of 128 (totaling 1.4 million parameters). This global attention stage is supervised using a Binary Cross-Entropy loss function. Training was conducted using the Lion optimizer \cite{lion} on four NVIDIA H100 GPUs with Distributed Data Parallel and a one-cycle learning rate schedule over 50 epochs within 12 hours.
The tracking stage employs a more compact configuration of the previously described setup, as it processes an average of only 55 hits per event following the filtering stage. The encoder and decoder of the tracking stage both use 2 layers with a hidden dimension of 32 amounting to an overall model size of 100 K parameters. Although Flash Attention is utilized, no windowing is applied due to the small number of hits remaining after filtering. Both multilayer perceptrons used for regression and object detection have two layers and twice the hidden dimension of the transformer. They are applied to the semantic representations of the queries after two Mask2Former decoder layers. Each detector hit is represented by a 22-dimensional feature vector, which includes: 
\begin{itemize}
    \item Global Cartesian coordinates ($x, y, z$) for both ends of the strip-like detector elements.
    \item Spatial covariance matrix components ($xx, xy, yx, yy$).
    \item Measured drift time and drift radius.
    \item Detector geometry identifiers: station, channel, layer, technology, and station $\phi/\eta$ indices.
    \item Augmented spatial features: the azimuthal angle $\phi$, the polar angle $\theta$, the radial distance from the beam axis ($R$), and the 3D distance from the interaction point.
\end{itemize} 
Training utilized a weighted multi-task loss function and was completed in about 8 hours running on a single NVIDIA H100 GPU. Hit-to-track assignment and object detection are supervised via Binary Cross-Entropy loss, using a parallel implementation of Hungarian matching for ground-truth assignment, while track parameter regression targets ($p_{\text{T}}$, $q$, $\eta$, and $\phi$) utilize Mean Squared Error. Although Dice and Focal losses were explored during ablation studies, they did not yield significant performance gains for the pattern recognition. 

Beyond architectural hyperparameter choices, a primary challenge in optimising performance was the selection of cost function weights for the Hungarian matching algorithm. The success of the training was found to be highly sensitive to these weights, with empirical results indicating that prioritizing the object detection task through a larger relative weight leads to improved performance. The code for the models explored in this work is available in the \href{https://github.com/samvanstroud/hepattn/tree/main/src/hepattn/experiments/atlas_muon}{\textbf{hepattn}} repository.

\subsection{ViT Results}
\begin{figure}[!htb]
  \centering
  \includegraphics[width=0.75\linewidth]{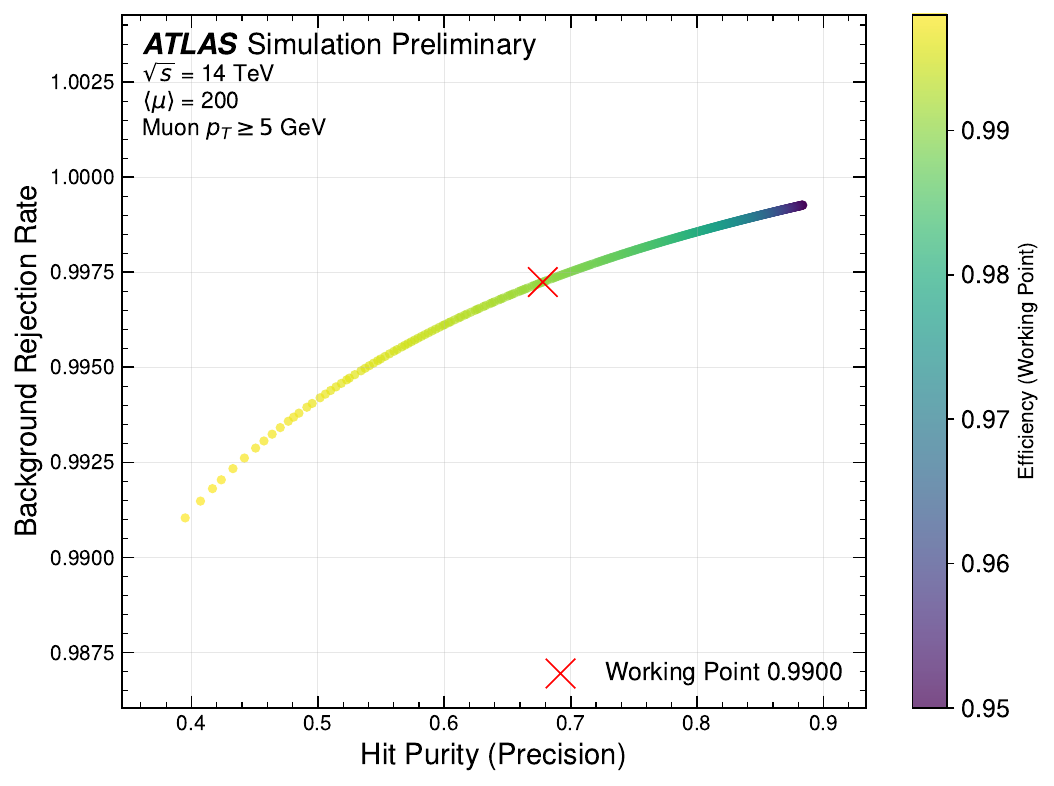}
  \caption{Background rejection rate as a function of hit purity (precision) for the hit-filtering stage \cite{ATLAS_MDET_2025}, evaluated on simulated $\sqrt{s} = 14$\,TeV events with $\langle\mu\rangle = 200$. Each point represents a different classification threshold, colored by the corresponding signal efficiency. The red marker indicates the 0.99 efficiency working point used to train the subsequent tracking stage.}
  \label{fig:filter_rejection_rate}
\end{figure}

\begin{figure}[!htb]
  \centering
  \includegraphics[width=\linewidth]{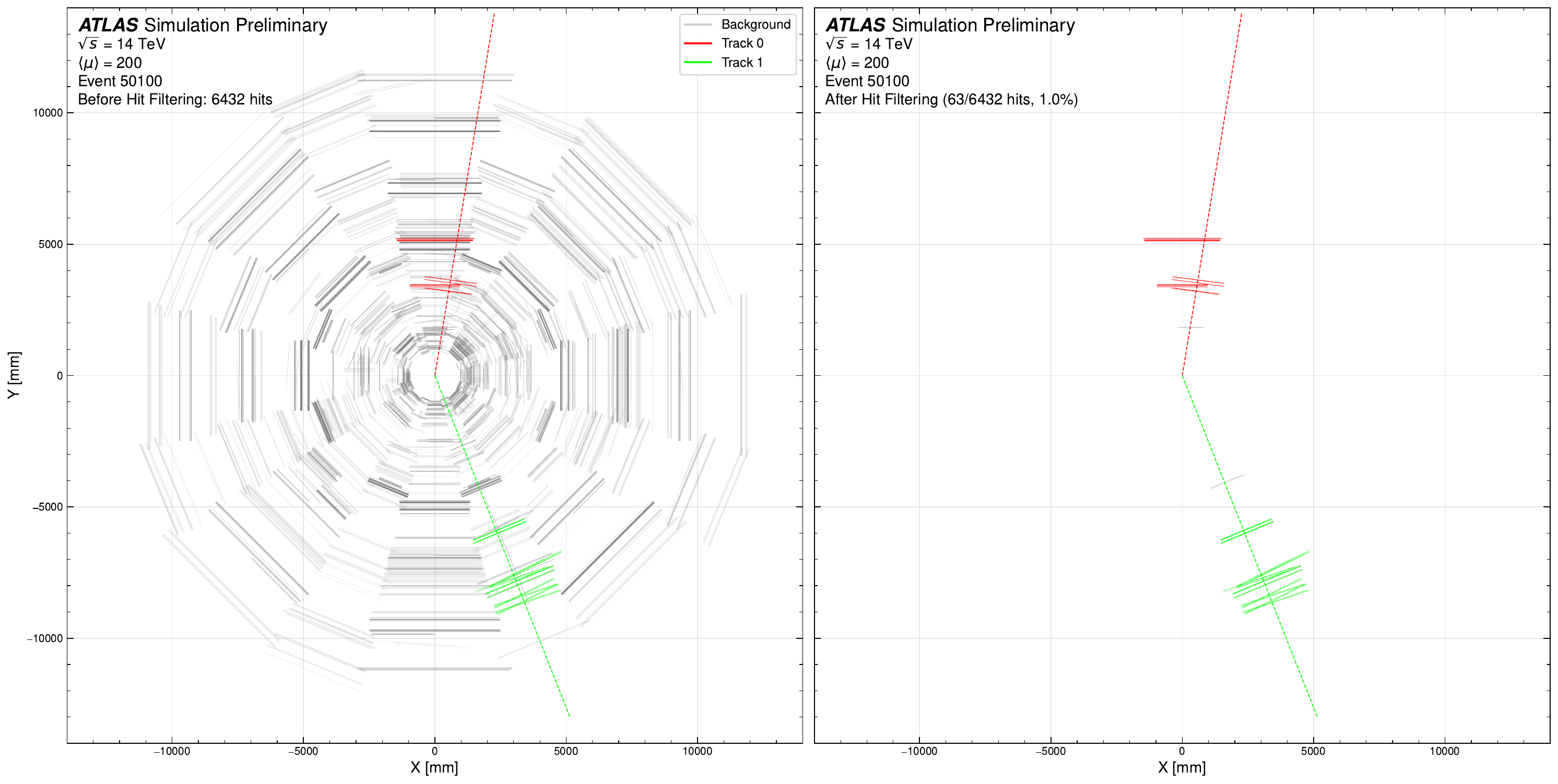}
  \caption{Event display in the ATLAS Muon Spectrometer at $\langle\mu\rangle=200$ \cite{ATLAS_MDET_2025}. The left panel shows the detector occupancy before hit filtering, while the right panel illustrates the hits remaining after the application of the background rejection stage, demonstrating a clear reduction in combinatorial noise. Hits are displayed using both ends of strip-like detector technology as straight lines. Hits belonging to muons from the pp-collision are color coded while the background hits are displayed in grey. All Barrels and EndCaps of the entire detector are being displayed.}
  \label{fig:event_display}
\end{figure}

The standalone performance of the hit-filtering stage highlights the effectiveness of the attention mechanism in capturing the heterogeneous global correlations of the Muon Spectrometer. The model achieves an Area Under the ROC Curve of 0.9997. This performance is quantified in Figure~\ref{fig:filter_rejection_rate}, which shows the background rejection rate as a function of hit purity (precision) across varying efficiency working points. The purity here is defined as the signal-to-noise ratio either before or after the hit filtering operation. On average, one to two muons recorded per event leave approx. 40 signal hits in the Muon Spectrometer. At a selected working point of 0.99 signal efficiency (indicated by the red marker), the hit purity increases from an initial value of approx. 0.6\% to approx. 66.5\%, corresponding to a background rejection rate of 99.7\%. In absolute terms, this reduces the average occupancy from 6,900 hits to 55 hits per event. Crucially, 99.7\% of the muon tracks remain reconstructable (defined as retaining at least three true hits) after the filtering stage. The impact of this reduction is visualized in the event display in Figure~\ref{fig:event_display}.

The inference speed of the model was evaluated on consumer-grade hardware to demonstrate its potential for affordable real-time applications. Measurements were performed on an NVIDIA RTX 3090 GPU with 24\,GB of vRAM (cost 1.5\,k CHF) and represent the time for a model forward pass once the data has been transferred to the GPU memory, thereby excluding input/output (IO) overhead. For a batch size of 200, the model achieves a processing time of 2.2\,ms per event with a GPU vRAM utilization of less than 80\%. When processing events individually (batch size 1), the execution time is 6.3\,ms per event with a vRAM utilization below 5\%. For a more modern GPU such as the NVIDIA H100 (cost 25\,k CHF), the inference time for a batch size of 200 decreases to 0.9\,ms per event at an average utilization of 20\%.

\begin{figure}[!htb]
  \centering
  \includegraphics[width=0.7\linewidth]{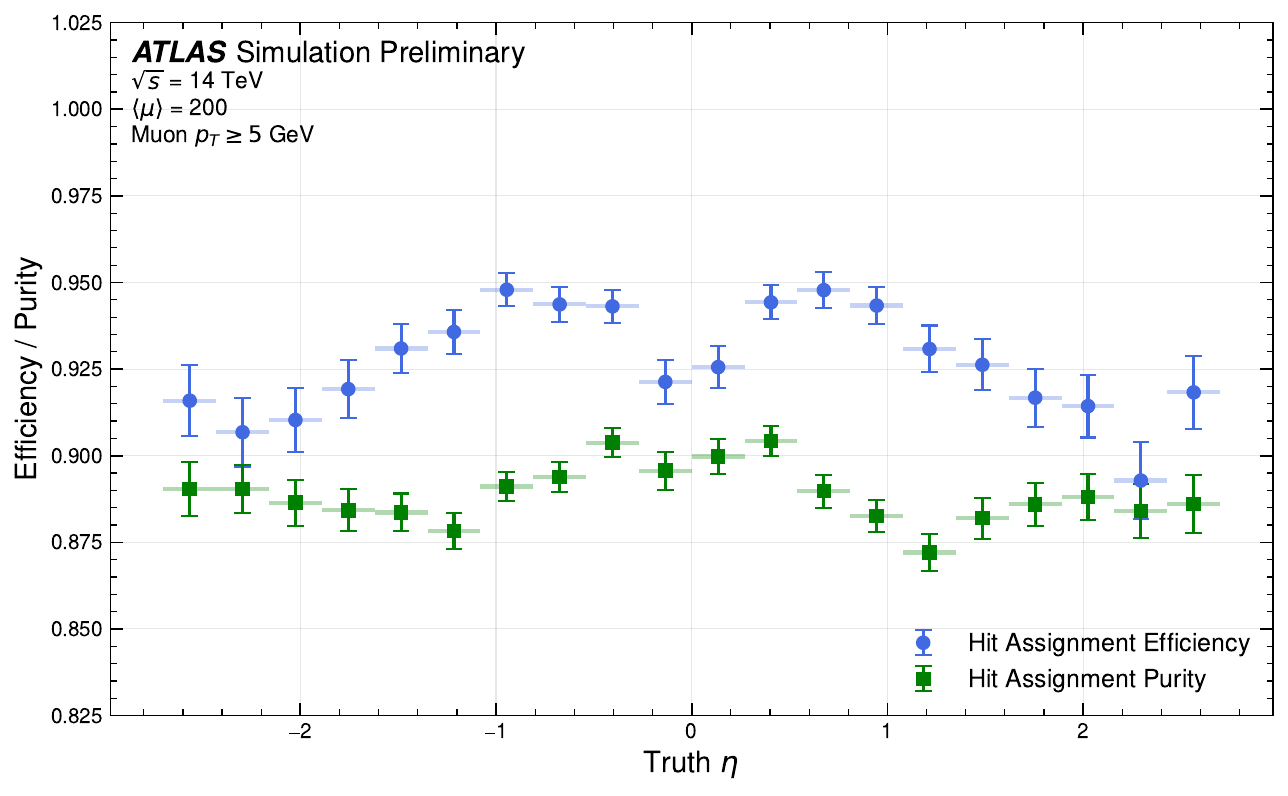}
  \caption{Hit assignment efficiency (blue) showing the average percentage of hits correctly assigned to each track as a function of pseudorapidity, and the corresponding purity (green) for the tracking stage \cite{ATLAS_MDET_2025}. This performance is evaluated on simulated proton-proton collisions at $\sqrt{s} = 14$\,TeV with $\langle\mu\rangle = 200$ using $t\bar{t}, J/\psi$ and $Z \rightarrow \mu\mu$ processes.}
  \label{fig:tracking_performance}
\end{figure}

The tracking stage achieves a signal detection efficiency of 98.0\% at a fake rate of 5.1\% in the ML-based reconstruction regime (meaning $p_{\text{T}} \ge 5$\,GeV and each track having more than three hits). The average hit assignment efficiency—representing the fraction of hits belonging to a muon that are correctly assigned to the reconstructed track—is 92.9\%, with a corresponding average hit assignment purity of 88.90\%. These results are summarized as a function of pseudorapidity in Figure~\ref{fig:tracking_performance}. 

\begin{figure}[!htb]
  \centering
  \includegraphics[width=0.7\linewidth]{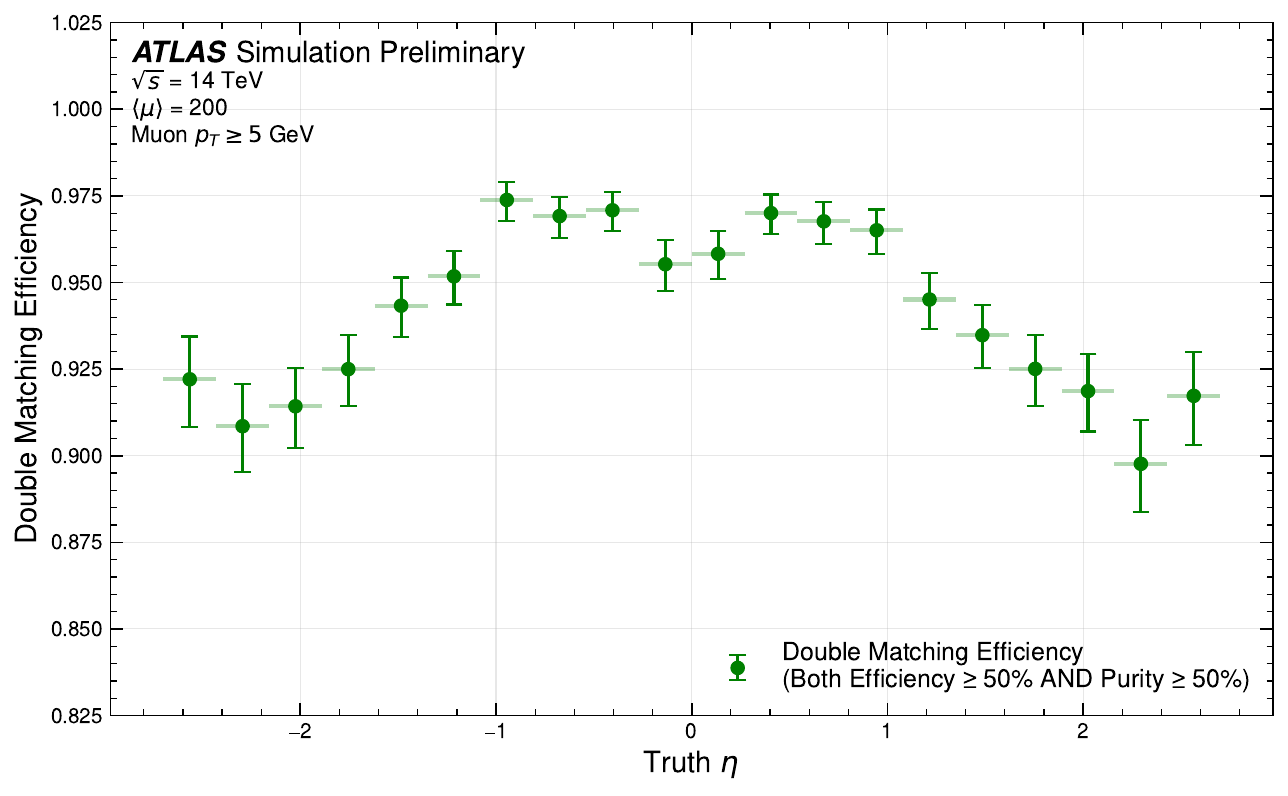}
  \caption{Double matching efficiency of the tracking stage as a function of true muon track pseudorapidity \cite{ATLAS_MDET_2025}. Double matching requires both efficiency and purity of hit-to-track assignment per track to be above 50\%. This performance is evaluated on simulated proton-proton collisions at $\sqrt{s} = 14$\,TeV with $\langle\mu\rangle = 200$ using $t\bar{t}, J/\psi$, and $Z \rightarrow \mu\mu$ processes.}
  \label{fig:double_matching}
\end{figure}

\begin{figure}[!htb]
  \centering
  \includegraphics[width=0.7\linewidth]{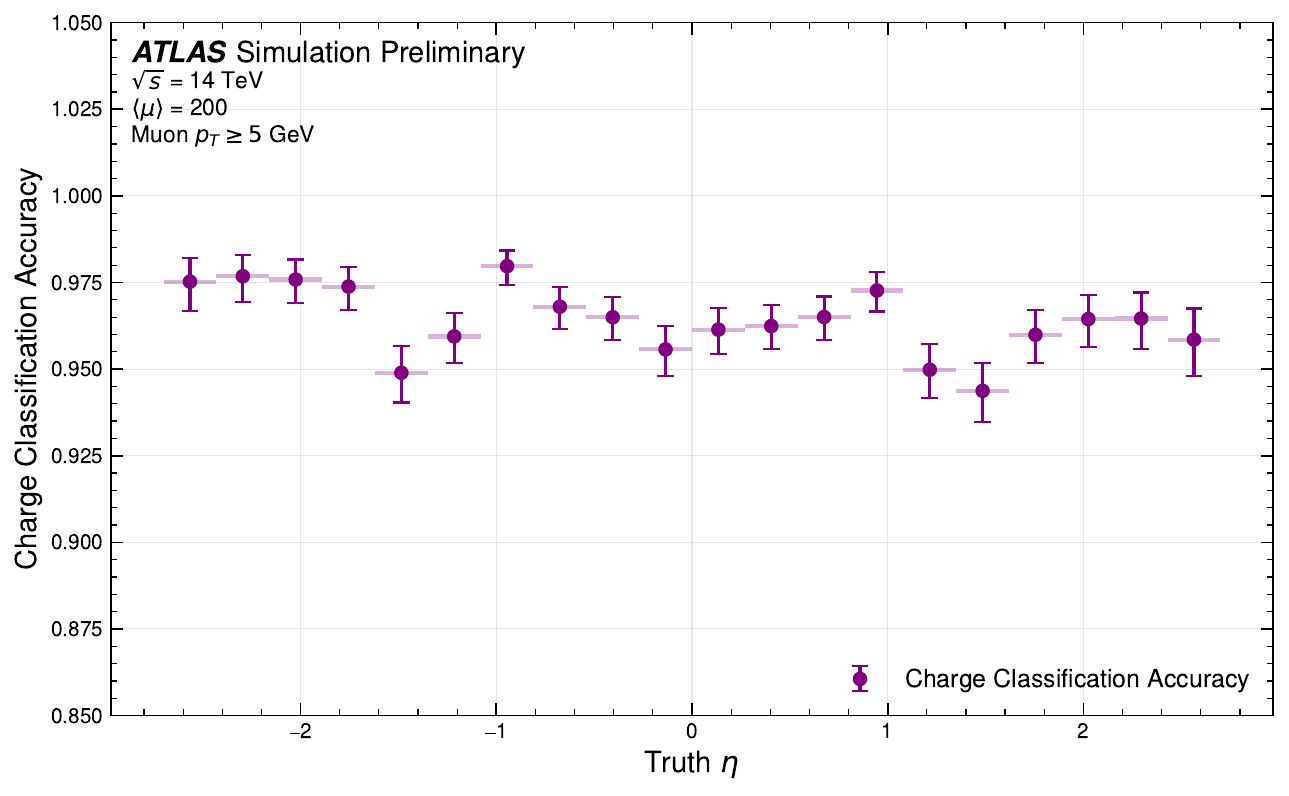}
  \caption{Charge sign classification accuracy of the tracking stage as a function of the true muon track pseudorapidity \cite{ATLAS_MDET_2025}. This evaluation uses simulated proton-proton collisions at $\sqrt{s} = 14$\,TeV with $\langle\mu\rangle = 200$ using $t\bar{t}, J/\psi$, and $Z \rightarrow \mu\mu$ processes.}
  \label{fig:charge_accuracy}
\end{figure}

To evaluate the overall quality of individual track reconstruction, we utilise the double matching efficiency. A track is considered double-matched if both its hit assignment efficiency and purity exceed a threshold of 50\%. As shown in Figure~\ref{fig:double_matching}, the model achieves an average double matching efficiency of 94.59\%. Furthermore, the model achieves a charge sign classification accuracy of 96.35\% on a balanced sample, as detailed in Figure~\ref{fig:charge_accuracy}. While the pattern recognition results are promising, the precision of track parameter regression is not yet competitive with the baseline reconstruction algorithm. The inference speed of this stage, again measured on an NVIDIA RTX 3090 excluding IO, is approx. 0.07\,ms per event for a batch size of 200 (vRAM utilization below 5\%), and 13.6\,ms per event for a batch size of 1 (vRAM utilization below 1\%).

Due to the compact architecture of the tracking model and the minimal memory footprint of the filtered hit data, the inference time for this stage is currently dominated by GPU kernel launch overheads rather than pure floating-point computation. This becomes evident considering that processing a single event requires nearly the same total wall-clock time as processing a batch of 200 events. As the model is not yet torch compiled, individual operations are executed as separate kernels without operator fusion. Future optimisations should focus on model compilation and graph-level fusion techniques to significantly reduce these overheads, mirroring the implementation of the global filtering stage where kernel fusion is already successfully implemented. The stark difference between the inference time and memory consumption of the global filtering stage and the tracking stage (2.2\,ms vs. 0.07\,ms) is primarily due to the fact that the tracking stage processes a significantly smaller number of hits (approx. 55 hits per event after filtering) compared to the global filtering stage (approx. 6,900 hits per event before filtering). In addition, the tracking stage's architecture results in about 14 times fewer parameters than the global filtering stage (100 K vs. 1.4 million parameters), which also contributes to the reduced memory footprint. 

\FloatBarrier

\section{Conclusions}

This work presented two machine learning approaches for muon reconstruction in the ATLAS Muon Spectrometer. The first method, utilizing Graph Neural Networks (GNNs) for background hit rejection, achieved a 15\% improvement in the reconstruction speed of the baseline algorithm in the Athena environment \cite{Athena}. Specifically, the GNN-based bucket filter reduced the per-event processing time from 255\,ms to 217\,ms when running on an NVIDIA H100 GPU, while introducing no negative bias into the baseline algorithm. This demonstrates the effectiveness of GNN-based filtering in managing the high-occupancy environments anticipated during the HL-LHC era.

The second approach, a proof-of-concept for end-to-end muon tracking using Vision Transformers (ViTs), leverages existing research from the computer vision community to solve the combinatorial problem of track finding. By adopting these state-of-the-art architectures, the tracking process does pattern finding and provides parameter estimates within 2.3\,ms on consumer-grade GPUs (cost approx. CHF 1.5 K). While the regression of precise track parameters and charge sign classification are not yet competitive with the baseline reconstruction, the model achieves a promising signal detection efficiency of 98.0\% and an average double matching efficiency of 94.59\%—satisfying the criterion of maintaining both hit assignment efficiency and purity above 50\% per track. These results demonstrate that the architecture provides not only high detection rates but also quality hit to track matching, as is also reflected in the average hit assignment efficiency and purity of 92.9\% and 88.90\%, respectively. Notably, the ViT-based approach exhibits a pattern-finding capability that is likely to be less reliant on specific sub-detector timing information compared to traditional algorithms considering the "broad ML-search regime" requiring $p_{\text{T}} \ge 5$\,GeV and 3 hits per track. In particular, ablation studies performed on augmented datasets without detector technologies such as RPCs and the New Small Wheel region showed similar performance in pattern recognition as the numbers presented in these proceedings. A quantitative comparison of hit-to-track assignment and object detection quality is unfortunately not yet possible at the time of publication, as the baseline algorithm is currently being adjusted for $\langle\mu\rangle=200$ conditions and official final benchmarks are not yet available.

Overall, this work demonstrates the promising potential of machine-learning-based methods for the ATLAS Muon Event Filter pipeline. For context, the baseline reconstruction takes 255\,ms per event on a single CPU thread (for a preliminary setup); however, the AMD EPYC 9654 processor can sustain approximately 200 concurrent threads, enabling parallel processing of many events simultaneously. The ViT-based pattern recognition achieves 2.3\,ms per event on a consumer-grade GPU, but this figure is amortised over a batch of 200 events processed in parallel. Both approaches therefore exploit parallelism at a comparable scale, making their effective per-event throughputs more comparable than the raw single-event latencies might suggest. Nevertheless, the roughly two-orders-of-magnitude reduction in per-event latency demonstrates how global attention mechanisms are well suited to solve complex combinatorial problems while fusing efficient attention implementations and state-of-the-art Computer Vision techniques. Furthermore, the attention mechanism is an algorithm that is likely to become even faster over the coming years; as the entire field of machine learning is heavily reliant on it, major industrial backing will likely ensure that attention-based architectures remain a highly sustainable technology choice for high-throughput HEP applications.

Future research could focus on evaluating the global filtering stage as a preprocessing step for the baseline reconstruction chain and exploring specialized configurations for long-lived particle (LLP) decays far from the interaction point in the Muon Spectrometer. Furthermore, there is potential to integrate hit-to-track assignment and initial parameter estimation directly into the ATLAS Event Filter to assist global track fits. Long-term development could also focus on enhancing robustness against detector aging and improving overall physics performance to ensure these methods become competitive with traditional algorithms in terms of track parameter precision. In addition, runtime optimizations such as pruning, quantization and model compilation could be explored to further reduce inference times and facilitate deployment. 

\Acknowledgements
This research was supported by the Eric and Wendy Schmidt Fund as part of the Next Generation Trigger Project at CERN.

\printbibliography


\end{document}

%% file: econfmacros.tex
\def\beq{\begin{equation}}
\def\eeq#1{\label{#1}\end{equation}}
\def\eeqn{\end{equation}}

\def\beqa{\begin{eqnarray}}
\def\eeqa#1{\label{#1}\end{eqnarray}}
\def\eeqan{\end{eqnarray}}

\let\bar=\overbar

\def\Dslash{\not{\hbox{\kern-4pt $D$}}}
\def\dslash{\not{\hbox{\kern-2pt $\del$}}}

\def\msb{{\bar{\ssstyle M \kern -1pt S}}}

